\documentclass[onecolumn,final]{elsart1p}

 \usepackage{graphicx}

\usepackage{amssymb}
\journal{Nuclear Physics A}

\begin{document}

\begin{frontmatter}



\title{A Microscopic Three-Cluster Model with Nuclear Polarization applied to the
Resonances of $^{7}Be$ and the Reaction $^{6}Li (  p,^{3}He ) ^{4}He$.}

\author[BITP,UA]{V.S. Vasilevsky} %
\author[UA]{F. Arickx} 
\ead{Frans.Arickx@ua.ac.be}
\author[UA]{J. Broeckhove}
\author[BITP]{T.P. Kovalenko}\\
 \address[BITP]{ Bogolyubov Institute for Theoretical Physics,
Kiev, Ukraine}
\address[UA]{Universiteit Antwerpen, Antwerpen, Belgium}

\begin{abstract}
A microscopic model for three-cluster configurations in light nuclei is
presented. It uses an expansion in terms of Faddeev components for which
the dynamic eqations are derived. The model
is designed to investigate binary channel processes in a compound system.
Gaussian and oscillator bases are used to expand the wave function and
to represent appropriate boundary conditions. We study the effect of cluster
polarization on ground and resonance states of $^{7}Be$, and
on the astrophysical $S$-factor of the reaction $^{6}Li(p,^{3}He)^{4}He$.

\end{abstract}

\end{frontmatter}

\section{Introduction}

Many of the light nuclei are weakly bound. Such nuclei can change their size
and shape considerably when interacting with other nuclei. We refer to this
phenomenon as cluster polarization. One expects cluster polarization to play
a role in reactions which involve light nuclei with small
separation energy such as the deuteron, $^{6}Li$,\ $^{7}Li$, and so on. One also expects
the effects to be more pronounced at small energies of the colliding nuclei, due to
the longer interaction time intervals.

Two different methods have been used to date to take into
account polarization of interacting clusters. The first, introduced by Tang et al.\
\cite{1982PhRvC..26.2410S,1988PhRvC..38.2013K,1982NuPhA.380...87K,
1982NuPhA.389..285K,1984PThPh..72..369K,1986NuPhA.457...93K},
considers internal monopole excitations to describe the polarization. The
second, introduced by the Kiev-Antwerp collaboration
\cite{kn:VVS+1984MonResE,Deumens84,kn:VVS+1984Reson8E,kn:cohstate2E,
kn:VVS+1990RGM+Sp2RE,kn:be8monop,kn:VV86collresE,2005PhRvC..71d4322S,2006JPhG...32.2137S},
is based on collective monopole and quadrupole polarizations of the compound nucleus.

In this paper we introduce a new approach in which we expand the three-cluster
many-particle wave function into Faddeev components. This approach
allows us to describe the proper boundary conditions for both binary and three-cluster
channels. We also introduce two different expansion schemes: a Gaussian basis to
describe bound two-cluster subsystems, and an oscillator basis to describe the relative
motion of the third cluster with respect to the two-cluster subsystem. The Gaussian basis
reproduces the intricate and complicated two-cluster bound-state behavior with a
limited number of terms, and is thus suited to describe cluster polarization.
The oscillator basis on the other hand allows for the proper representation of the
scattering boundary condition in the matrix form of the Schr\"odinger equation.
We derive a set of equations for the Faddeev components within the Coupled Channels
Formalism.

We apply this approach to cluster polarization in $^{7}Be$. This system
exhibits a well determined set of bound and resonance states, and has
been thoroughly studied by many microscopical methods
\cite{2002NuPhA.699..963A,2001PhRvC..63d4611A,1988PhRvC..38.1531F,
1985PhRvC..31..342F,1984NuPhA.419..133W,1986NuPhA.459..387M,
1983NuPhA.410..208H,1983PhRvC..28...57W,1986NuPhA.460..559K,1984NuPhA.413..323K}.
Moreover, two reactions which are 
connected to this nucleus, $^{3}He\left(  \alpha,\gamma\right)  ^{7}Be$ and
$^{6}Li\left(  p,^{3}He\right)  ^{4}He$, are important in astrophysical
models \cite{kn:adelberger98,2005NuPhA.752..522C}. The former
reaction has been extensively investigated by different microscopic and
semi-microscopic methods and is involved in the solar neutrino problem
\cite{1983PhRvC..28...57W,1984NuPhA.419..133W,1986NuPhA.460..559K,1986NuPhA.459..387M, 2000FBS....29..121C,PhysRevC.48.1420,1988JPhG...14L.211B}. The
latter reaction is connected to the big-bang nucleosynthesis and determines the abundance
of light elements in the universe. It has received much less attention in the literature.
It was investigated within a three-cluster microscopic model in an astrophysically
relevant energy range \cite{2002NuPhA.699..963A}, and also in a multi-configuration resonating
group model \cite{1988PhRvC..38.1531F} for a wide energy range.

We model the $^{7}Be$ nucleus using a many-channel cluster wave function containing
both two-cluster and three-cluster components. As we wish to consider both the two-cluster
components $^{4}He+^{3}He$ and $^{6}Li +p$, we use the $^{4}He+d+p$ three-cluster
configuration.

In the reaction $^{6}Li\left( p,^{3}He\right) ^{4}He$ one structureless subsystem,
the proton, and three cluster subsystems $^{6}Li$, $^{4}He$ and $^{3}He$ are involved.
They are connected
with the lowest binary channels $^{4}He+^{3}He$ and $^{6}Li +p$ which define
the main properties of the bound and some resonance states of $^{7}Be$.

Only 1.5 MeV is necessary to split the $^{6}Li$ nucleus in
$^{4}He$ and $d$. To disintegrate the $^{3}He$ nucleus in a deuteron and a
proton the total energy of $^{3}He$ has to exceed 5.5 MeV.  To split the $^{4}He$
nucleus into a proton and $^{3}H$ already more than 20 MeV is needed.
This leads one to expect that cluster polarization of the $^{6}Li$ and $^{3}He$
nuclei will be important and should be taken into account for the low-energy
states of $^{7}Be$, and that the polarization of $^{4}He$ can be neglected.

The new method presented in this paper achieves two goals: (1) it allows to study
the polarizability of weakly bound two-cluster systems induced by an incident cluster,
and (2) it provides a description of the resonance structure of the resulting
three-cluster system.

\section{The three-cluster model}

\subsection{Model space and Hamiltonian}

We introduce our approach for the general case of $s$-shell clusters, but it can in principle be
extended to cover clusters with an arbitrary number of nucleons. It is analogous to the one
formulated in \cite{A4Varenna2003} and \cite{A4Resonances2004}. However, in those contributions
a bi-oscillator basis was used to study the three-cluster interaction, while we will include
both Gaussian and oscillator basis states.

A Gaussian basis is a multi-parameter variational basis that can reproduce
complicated and intricate inter-cluster wave functions with few terms, thus achieving
high numerical precision with low computational complexity. It has been considered
on different occasions in microscopic calculations, and found to be
very efficient for bound states, even for loosely bound nuclei with proton
and neutron excess \cite{Kukulin:1977ux,1994NuPhA.571..447V,1996PhRvA..53.1907V,
1997NuPhA.616..383V,Varga:1995dm,2003PrPNP..51..223H}. Its drawback is the
non-orthogonality of the basis functions that can lead to numerical instabilities.
We will adopt this basis to represent the (weakly) bound two-cluster subsystems
in the three-cluster description.

The oscillator basis is suitable for the description of bound as well as scattering boundary
conditions \cite{kn:Fil_Okhr,kn:Fil81,kn:Heller1,kn:Heller2,kn:Yamani},
and the corresponding matrix form of the Schr\"{o}dinger equation
\cite{1982JMP....23...83Y,2000AnPhy.280..299B} is
similar to the $R$-matrix theory for nuclear reactions. Due to the orthogonality of the basis
functions it does not suffer from numerical instabilities but it converges
more slowly than the Gaussian basis. It was shown that an
acceptable precision for light $p$-shell
nuclei can be achieved with 30 to 50 oscillator functions 
\cite{kn:cohstate2E,kn:VV86collresE,kn:VR33_42}.
In some model situations this number can be further reduced, even down to 3 or 5 functions, as
was shown in \cite{kn:VA_PR}.
The calculation of matrix elements of different operators between oscillator
functions can be done with the technique of the Generalized Coherent States
\cite{kn:cohstate1E,kn:cohstate2E,kn:3cl-theory-PhRevC-2001-63-034606}, which leads to recurrence
relations for the matrix elements. We will consider this basis to describe
the scattering component in the three-cluster system.

The wave function for $s$-shell clusters can be written as
\begin{equation}
\Psi^{J}=\widehat{\mathcal{A}}\left\{
  \left[ \Phi_{1}\left( A_{1}\right)
         \Phi_{2}\left( A_{2}\right)
         \Phi_{3}\left( A_{3}\right) \right]^{S}
  \left[ f_{1}^{L}\left( \mathbf{x}_{1},\mathbf{y}_{1}\right)
       + f_{2}^{L}\left( \mathbf{x}_{2},\mathbf{y}_{2}\right)
       + f_{3}^{L}\left( \mathbf{x}_{3},\mathbf{y}_{3}\right) \right]
  \right\} ^{J} \label{eq:001}%
\end{equation}
where $\Phi_{\alpha}\left(  A_{\alpha}\right)  $ is a shell-model wave
function for the internal motion of cluster $\alpha$ ($\alpha=1,2,3$) and
$f_{\alpha}^{L}\left(  \mathbf{x}_{\alpha},\mathbf{y}_{\alpha}\right)  $ is a
Faddeev component. The first factor describes the internal cluster motion and has total
orbital angular momentum $L=0$ because of the $s$-shell clusters, and only its total
spin quantum number is indicated.
The second factor represents the relative inter-cluster motion, and is responsible for
the total orbital angular momentum $L$. We consider an $LS$ coupling scheme so that
$L$ and $S$ couple to the total angular momentum $J$.

It is well know that Faddeev components are very suitable for implementing the
necessary boundary conditions for binary as well as for three-cluster channels
\cite{kn:faddeev+merk93}.

In the Faddeev component $f_{\alpha}^{L}\left(  \mathbf{x}_{\alpha},\mathbf{y}_{\alpha}\right)$,
$\mathbf{x}_{\alpha}$ is the Jacobi vector proportional to
the distance between the $\beta$ and $\gamma$ clusters ($\alpha,\beta$ and $\gamma$ form a
cyclic permutation of 1, 2 and 3), while $\mathbf{y}_{\alpha}$ is the Jacobi vector connecting
the $\alpha$ cluster to the center of mass of
the $\beta$\ and $\gamma$ clusters:
\begin{eqnarray}
\label{eq:002}%
\mathbf{x}_{\alpha}  &  =& \sqrt{\frac{A_{\beta}A_{\gamma}}{A_{\beta}+A_{\gamma
}}}\left(  \frac{1}{A_{\beta}}\sum_{j\in A_{\beta}}\mathbf{r}_{j}-\frac
{1}{A_{\gamma}}\sum_{k\in A_{\gamma}}\mathbf{r}_{k}\right)  \qquad
\label{eq:002a}\\
\mathbf{y}_{\alpha}  &  =& \sqrt{\frac{A_{\alpha}\left(  A_{\beta}+A_{\gamma
}\right)  }{A_{\alpha}+A_{\beta}+A_{\gamma}}}\left(  \frac{1}{A_{\alpha}}%
\sum_{i\in A_{\alpha}}\mathbf{r}_{i}-\frac{1}{A_{\beta}+A_{\gamma}}\left[
\sum_{j\in A_{\beta}}\mathbf{r}_{j}+\sum_{k\in A_{\gamma}}\mathbf{r}%
_{k}\right]  \right)  \label{eq:002b}%
\end{eqnarray}

For each Faddeev component we use bi-spherical harmonics
\begin{equation}
f_{\alpha}^{L}\left(  \mathbf{x}_{\alpha},\mathbf{y}_{\alpha
}\right)  =\sum_{\lambda_{\alpha},l_{\alpha}}f_{\alpha}^{\left(
\lambda_{\alpha},l_{\alpha};L\right)  }\left(  x_{\alpha},y_{\alpha}\right)
\left\{  Y_{\lambda_{\alpha}}\left(  \widehat{\mathbf{x}}_{\alpha}\right)
Y_{l_{\alpha}}\left(  \widehat{\mathbf{y}}_{\alpha}\right)  \right\}  _{LM}
\label{eq:003}%
\end{equation}
which lead to the four quantum numbers $\lambda_{\alpha},l_{\alpha},LM$.  
The parity of the three-cluster states
is then determined by the partial angular momenta: $\pi=\left(  -\right)
^{\lambda_{\alpha}+l_{\alpha}}$.

The radial part of the Faddeev components $f_{\alpha}^{\left(
\lambda_{\alpha},l_{\alpha};L\right)}$ is obtained by using products of
Gaussian basis functions $\left\{
G_{\lambda_{\alpha}}(\mathbf{x}_{\alpha},b_{\nu_{\alpha}})\right\}  $ and
oscillator basis functions $\left\{  \Phi_{n_{\alpha}l_{\alpha}%
}(\mathbf{y}_{\alpha},b)\right\}$, where
\begin{eqnarray}
\Phi_{nl}(\mathbf{y},b)
  & = & (-1)^{n} \frac{1}{b^{3/2}} N_{nl}\rho^{l}
      L_{n}^{l+1/2}\left(  \rho^{2}\right) \exp\left(  -\rho^{2}/2\right)
      Y_{lm}\left(  \widehat{\mathbf{y}}\right) \label{eq:004} \\
  & = & \Phi_{nl}\left( y,b \right) Y_{lm}\left( \widehat{\mathbf{y}}\right)
      \qquad\left( \rho  =\frac{y}{b},\ \: N_{nl}=\sqrt{\frac{2\ \Gamma\left(  n+1\right)
}{\Gamma\left(  n+l+3/2\right)  }} \right) \nonumber
\end{eqnarray}
represents an oscillator function and
\begin{eqnarray}
G_{\lambda}\left(  \mathbf{x},b_{\nu}\right)
  & =& \frac{1}{b_{\nu}^{3/2}}
     \sqrt{\frac{2}{\Gamma\left( \lambda+3/2\right) }}
     \rho^{\lambda}\exp\left\{-\frac{1}{2}\rho^{2}\right\}
     Y_{\lambda\mu}\left(  \widehat{\mathbf{x}}\right)
   \label{eq:005}   \\
  & =& G_{\lambda}\left(  x,b_{\nu}\right)
      Y_{\lambda\mu}\left( \widehat{\mathbf{x}}\right)
      \qquad\left( \rho =\frac{x}{b_{\nu}} \right) \nonumber
\end{eqnarray}
stands for a Gaussian function. One then immediately obtains the radial part
\begin{equation}
f_{\alpha}^{\left(\lambda_{\alpha},l_{\alpha};L\right) }\left( x_{\alpha},y_{\alpha}\right) =
  \sum_{\nu_{\alpha}, n_{\alpha}}
  C_{\nu_{\alpha}\lambda_{\alpha};n_{\alpha}l_{\alpha}}^{\left(\alpha\right)}
                   G_{\lambda_{\alpha}}\left(  x_{\alpha},b_{\nu_{\alpha}}\right)
                   \Phi_{n_{\alpha}l_{\alpha}}\left( y_{\alpha},b \right) 
\label{eq:006}
\end{equation}

Introducing the Gaussian-Oscillator bi-spherical expansion in the total wave function
$\Psi^{J}$ in (\ref{eq:001}) leads to the form
\begin{eqnarray}
 & \Psi^{J}
& =  \sum_{\alpha}\sum_{\lambda_{\alpha},l_{\alpha}}\sum_{\nu_{\alpha
},n_{\alpha}}C_{\nu_{\alpha}\lambda_{\alpha};n_{\alpha}l_{\alpha}}^{\left(
\alpha\right)  }\nonumber\\
&  \times & \widehat{\mathcal{A}}\left\{  \left[  \Phi_{1}\left(  A_{1}\right)
\Phi_{2}\left(  A_{2}\right)  \Phi_{3}\left(  A_{3}\right)  \right]
^{S}\left[  G_{\lambda_{\alpha}}(\mathbf{x}_{\alpha},b_{\nu_{\alpha}}%
)\Phi_{n_{\alpha}l_{\alpha}}(\mathbf{y}_{\alpha},b)\right]  ^{L}\right\}  ^{J} \nonumber \\
& = & \sum_{\alpha}\sum_{\lambda_{\alpha},l_{\alpha}}\sum_{\nu_{\alpha
},n_{\alpha}}C_{\nu_{\alpha}\lambda_{\alpha};n_{\alpha}l_{\alpha}}^{\left(
\alpha\right)  }  \\
& \times & \widehat{\mathcal{A}}\left\{  \left[  \Phi_{1}\left(
A_{1}\right)  \Phi_{2}\left(  A_{2}\right)  \Phi_{3}\left(  A_{3}\right)
\right]  _{S}G_{\lambda_{\alpha}}(x_{\alpha},b_{\nu_{\alpha}})\Phi_{n_{\alpha
}l_{\alpha}}(y_{\alpha},b)\left\{  Y_{\lambda_{\alpha}}\left(  \widehat
{\mathbf{x}}_{\alpha}\right)  Y_{l_{\alpha}}\left(  \widehat{\mathbf{y}%
}_{\alpha}\right)  \right\}  _{L}\right\}  _{J} \nonumber \label{eq:003b}%
\end{eqnarray}
and will subsequently be referred to using the acronym GOB.

The microscopic hamiltonian for a three-cluster configuration can be written as
\begin{equation}
\widehat{H}=\widehat{T}+\widehat{V}=\sum_{\alpha=1}^{3}\widehat{H}_{\alpha
}^{\left(  1\right)  }+\widehat{T}_{r}+\sum_{\alpha}\widehat{V}_{\alpha}
\label{eq:006b}%
\end{equation}
i.e.\ a sum of three single-cluster hamiltonians $\widehat{H}_{\alpha
}^{\left(  1\right)  }$ describing the internal structure of each cluster, and
a term responsible for the inter-cluster dynamics. The latter consists of the
kinetic energy operator for relative motion of clusters $\widehat{T}_{r}$ and
the potential energy of the interaction between clusters. This hamiltonian can be
also expressed as a sum of one two-cluster hamiltonian, and terms
representing the interaction of the third cluster with the two-cluster subsystem:
\begin{equation}
\widehat{H}=\widehat{H}_{\alpha}^{\left(  2\right)  }+\widehat{H}_{\alpha
}^{\left(  1\right)  }+\widehat{T}_{\alpha}+\sum_{\beta\neq\alpha}\widehat
{V}_{\beta}. \label{eq:006c}%
\end{equation}
The terms appearing in (\ref{eq:006b}) and (\ref{eq:006c}) are easily expanded
in terms of the particle operators:
\begin{eqnarray}
\widehat{T}_{r} 
    & = & \frac{\hbar^{2}}{2m}\Delta_{\mathbf{x}_{\alpha}}%
      +\frac{\hbar^{2}}{2m}\Delta_{\mathbf{y}_{\alpha}} \label{eq:007a} \\
\widehat{T}_{\alpha}
     & = & \frac{\hbar^{2}}{2m}\Delta_{\mathbf{y}_{\alpha}} \\
\widehat{H}_{\alpha}^{\left(  1\right)  }
     & = & \sum_{i\in A_{\alpha}} \widehat{T}\left(  i\right)
      +\sum_{i<j\in A_{\alpha}}\widehat{V}\left(ij\right) \label{eq:007b} \\
\widehat{H}_{\alpha}^{\left(  2\right)  }
     & = & \sum_{i\in A_{\beta}+A_{\gamma}}\widehat{T}\left(  i\right)
      +\sum_{i<j\in A_{\beta}+A_{\gamma}}\widehat{V}\left(  ij\right) \label{eq:007c} \\
\widehat{V}_{\alpha}
     & = & \sum_{i\in A_{\beta}}\sum_{j\in A_{\gamma}} \widehat{V}\left(ij\right)
\label{eq:007d}
\end{eqnarray}

The wave function of a two-cluster subsystem
\begin{equation}
\psi_{\alpha}^{J_{\alpha}\lambda_{\alpha}}=\widehat{\mathcal{A}}\left\{
\left[  \Phi_{\beta}\left(  A_{\beta}\right)  \Phi_{\gamma}\left(  A_{\gamma
}\right)  \right]  ^{S_{\alpha}}\phi_{\lambda_{\alpha}}(x_{\alpha}%
)Y_{\lambda_{\alpha}}\left(  \widehat{\mathbf{x}}_{\alpha}\right)  \right\}
^{J_{\alpha}}
\end{equation}
is expanded in Gaussian cluster functions
\begin{equation}
\psi_{\alpha}^{J_{\alpha}\lambda_{\alpha}}=\sum_{\nu}D_{\lambda_{\alpha},\nu}^{\left(
\alpha\right)  }\chi_{\nu;\alpha}^{J_{\alpha}\lambda_{\alpha}},
\end{equation}
where
\begin{equation}
\chi_{\nu;\alpha}^{J_{\alpha}\lambda_{\alpha}}=\widehat{\mathcal{A}}\left\{
\left[  \Phi_{\beta}\left(  A_{\beta}\right)  \Phi_{\gamma}\left(  A_{\gamma
}\right)  \right]  ^{S_{\alpha}}G_{\lambda_{\alpha}}(x_{\alpha},b_{\nu
})Y_{\lambda_{\alpha}}\left(  \widehat{\mathbf{x}}_{\alpha}\right)  \right\}
^{J_{\alpha}}. \label{eq:011}
\end{equation}

The bound states $E_{\sigma}^{\left( \alpha\right) }$ of this subsystem ($\sigma=0$ is
the ground state, $\sigma >0$ are excited or pseudo-bound states), and their corresponding
eigenstates $\phi_{\lambda_{\alpha} }^{\left( \alpha, \sigma\right)  }$
are defined by $\left\{ D^{(\alpha,\sigma)}_{ \lambda_{\alpha}, \nu} \right\}$ and can be
obtained by solving the corresponding generalized eigenvalue problem
\begin{equation}
\sum_{\tilde{\nu}=1}^{N_{\alpha}}
\left\langle \nu,\alpha
\left\vert \widehat{H}_{\alpha}^{\left(2\right)}-E_{\sigma}^{\left( \alpha\right)}\right\vert \tilde{\nu},\alpha\right\rangle D^{(\alpha,\sigma)}_{ \lambda_{\alpha}, \tilde{\nu}}  =0
\label{eq:010}
\end{equation}
The number of terms in (\ref{eq:011}), and correspondingly the number of eigenstates,
is chosen for sufficient convergence of the ground state, and depends on the two-cluster
subsystem labeled by $\alpha$.
The corresponding wave functions for the two-cluster relative motion are
\begin{equation}
\sum_{\nu_{\alpha}=1}^{N_{\alpha}^{\left(  G\right)  }}D_{\lambda_{\alpha},\nu_{\alpha}
}^{\left(  \sigma,\alpha\right)  }G_{\lambda_{\alpha}}(x_{\alpha}
,b_{\nu_{\alpha}})=\phi_{\lambda_{\alpha}}^{\left(\alpha,  \sigma\right)
}\left(  x_{\alpha}\right)
\end{equation}

Because of the Pauli principle between nucleons, it is hard to unambiguously
derive a set of Faddeev type equations for the Faddeev three-cluster amplitudes
$f_{\alpha}(\mathbf{x}_{\alpha},\mathbf{y}_{\alpha})$ in a fully microscopic
three-cluster description. An attempt to achieve this
has recently been proposed in \cite{kn:JPhysConfSer-111-2055}.
In the current paper we solve the Schr\"odinger equation through the traditional
coupled channels formalism \cite{Tobocman61,Austern70} to obtain the
$f_{\alpha}(\mathbf{x}_{\alpha},\mathbf{y}_{\alpha})$ amplitudes.

The dynamic equations for the three-cluster system are easily obtained by
substituting (\ref{eq:003b}) in the the Schr\"odinger equation containing the
Hamiltonian (\ref{eq:006b}), and in order to solve for the expansion coefficients 
$C_{\nu_{\alpha}\lambda_{\alpha};n_{\alpha}l_{\alpha}}^{\left( \alpha\right)  }$,
appropriate boundary conditions have to be expressed in terms of the expansion basis.
This is done in the next section, exploiting the physical relevance of the
two-cluster eigenstates discussed above.

\subsection{Boundary conditions}

In this paper we focus on the energy range between the ground state of $^{7}Be$
and the three-cluster threshold for $^{4}He+d+p$ disintegration.
We therefore only have to consider binary scattering and reaction channels, and
can neglect three-cluster decay. Thus only two-cluster asymptotics need to be
included in the boundary conditions. In this case $x_{\alpha
}\ll y_{\alpha}$, i.e.\ one cluster is at a large distance of the other two
clusters, and the latter will constitute a bound two-cluster subsystem.

For large values of the Jacobi vector $y_{\alpha}$, the
function $f_{\alpha}^{\left(  \lambda_{\alpha},l_{\alpha};L\right)  }\left(
x_{\alpha},y_{\alpha}\right)$ asymptotically factorizes as
\begin{equation}
f_{\alpha}^{\left(  \lambda_{\alpha},l_{\alpha};L\right)  }\left(  x_{\alpha
},y_{\alpha}\right)  \approx\phi_{\lambda_{\alpha}}^{\left(
\alpha,\sigma\right)  }\left(  x_{\alpha}\right)  \left[  S_{c_{0},c_{\alpha}}%
\psi_{l_{\alpha}}^{\left(  -\right)  }\left(  p_{\alpha}y_{\alpha}\right)
-S_{c_{0},c_{\alpha}}\psi_{l_{\alpha}}^{\left(  +\right)  }\left(  p_{\alpha
}y_{\alpha}\right)  \right]  \label{eq:101}%
\end{equation}
for continuum states and
\begin{equation}
f_{\alpha}^{\left(  \lambda_{\alpha},l_{\alpha};L\right)  }\left(  x_{\alpha
},y_{\alpha}\right)  \approx-\phi_{\lambda_{\alpha}}^{\left(
\alpha,\sigma\right)  }\left(  x_{\alpha}\right)  \left[  S_{c_{0},c_{\alpha}}%
\psi_{l_{\alpha}}^{\left(  +\right)  }\left(  -i\left\vert p_{\alpha
}\right\vert y_{\alpha}\right)  \right]  \label{eq:102}%
\end{equation}
for bound states. 
The entrance channel is denoted by $c_{0}$, and $c_{\alpha}$ refers to the current
channel where $\alpha$ stands short for all necessary ($\lambda_{\alpha},l_{\alpha},\ldots$)
quantum numbers.
The momentum $p_{\alpha}$ is
defined by
\begin{equation}
p_{\alpha}=\sqrt{\frac{2m}{\hbar^{2}}\left(  E-E_{\sigma}^{\left(
\alpha\right)  }\right)  } \label{eq:103}%
\end{equation}
and the bound state energy $E_{\sigma}^{\left(  \alpha\right)  }$ of the
two-cluster subsystem determines the threshold energy of the $c_{\alpha}$ channel.

This factorization of the wave function (\ref{eq:003b}) also occurs in the expansion
coefficients
$\left\{ C_{\nu_{\alpha}\lambda_{\alpha};n_{\alpha}l_{\alpha}}^{\left(\alpha\right)}\right\}$.
The asymptotic region in this representation is connected to large values of $n_{\alpha}$
(for more details see \cite{kn:Fil_Okhr,kn:Fil81,kn:VA_PR,kn:cohstate2E}), and there the
coefficients factorize as:
\begin{eqnarray}
C_{\nu_{\alpha}\lambda_{\alpha};n_{\alpha}l_{\alpha}}^{\left(  \alpha\right)}  
&  \approx  &D^{(\alpha,\sigma)}_{ \lambda_{\alpha}, \nu_{\alpha}}
C_{n_{\alpha}l_{\alpha}}^{\left(c_{\alpha}\right)  } \nonumber\\
&  = &D^{(\alpha,\sigma)}_{ \lambda_{\alpha}, \nu_{\alpha}}%
\sqrt{2r_{n_{\alpha}}}\left[  S_{c_{0},c_{\alpha}}\psi_{l_{\alpha}}^{\left(
-\right)  }\left(  p_{\alpha}r_{n_{\alpha}}\right)  -S_{c_{0},c_{\alpha}}%
\psi_{l_{\alpha}}^{\left(  +\right)  }\left(  p_{\alpha}r_{n_{\alpha}}\right)
\right]  \label{eq:104} \\
C_{\nu_{\alpha}\lambda_{\alpha};n_{\alpha}l_{\alpha}}^{\left(  \alpha\right)}
& \approx & D^{(\alpha,\sigma)}_{ \lambda_{\alpha}, \nu_{\alpha}}
C_{n_{\alpha}l_{\alpha}}^{\left(  c_{\alpha}\right)}=
-D^{(\alpha,\sigma)}_{ \lambda_{\alpha}, \nu_{\alpha}}\sqrt{2r_{n_{\alpha}}}
\left[S_{c_{0},c_{\alpha}}\psi_{l_{\alpha}}^{\left(  +\right)  }
\left(  -i\left\vert p_{\alpha}\right\vert r_{n_{\alpha}}\right)  \right]
\label{eq:104a}
\end{eqnarray}
where
\begin{equation}
r_{n_{\alpha}}=b\sqrt{4n_{\alpha}+2l_{\alpha}+3}, \label{eq:105}%
\end{equation}
is the classical oscillator turning point corresponding to the oscillator length $b$,
and $\psi_{l_{\alpha}}^{\left( -\right)}\left( p_{\alpha}r_{n_{\alpha}}\right)$
(respectively
$\psi_{l_{\alpha}}^{\left(+\right)  }\left(  p_{\alpha}r_{n_{\alpha}}\right)$)
are the familiar radial Coulomb modified incoming (respectively outgoing) wave
functions, normalized to unit flux (see for instance \cite{1983RvMP...55..155B}).

The equations (\ref{eq:104}) and (\ref{eq:104a}) represent the boundary
condition for the expansion coefficients
$\left\{ C_{\nu_{\alpha}\lambda_{\alpha};n_{\alpha}l_{\alpha}}^{\left( \alpha\right) }\right\}$
for scattering and bound states in the $c_{\alpha}$ binary channel.

\subsection{The Dynamic equations}

The many-channel equations of the GOB model can be solved in three stages.

In the first step the Schr\"{o}dinger equation for all two-cluster
subsystems is solved. This is done by diagonalizing the $N_{\nu
}\times N_{\nu}$ matrix of the two-cluster hamiltonian
\begin{equation}
\left\Vert \left\langle \nu_{\alpha},\lambda_{\alpha}\left\vert \widehat
{H}_{\alpha}^{\left(  2\right)  }\right\vert \tilde{\nu}_{\alpha}
,\lambda_{\alpha}\right\rangle \right\Vert
\end{equation}
between the cluster Gaussian functions of (\ref{eq:011}). The discrete set of
eigenvalues $E_{\sigma}^{\left(\alpha\right)}$ correspond to bound states, or
to pseudo-bound states above the threshold that are artifacts of the
diagonalization in a finite basis. The eigenstate wave function is
$\left\{ D_{\lambda_{\alpha},\nu_{\alpha}}^{\left(\sigma_{\alpha}, \alpha \right)}\right\}$.
This step has to be repeated for every value of the partial angular momentum
$\lambda_{\alpha}$ considered in the full calculation.

In the second step the block matrix of the total three-cluster hamiltonian
\begin{equation}
\left\Vert \left\langle
\nu_{\alpha},\lambda_{\alpha};n_{\alpha},l_{\alpha}
\left\vert \widehat{H} \right\vert
\nu_{\tilde{\alpha} },\lambda_{\tilde{\alpha}};n_{\tilde{\alpha}},l_{\tilde{\alpha}}
\right\rangle \right\Vert
\end{equation}
is transformed to the representation of two interacting clusters using the aforementioned
eigenfunctions. One obtains
\begin{equation}
\left\Vert \left\langle
\sigma_{\alpha},\lambda_{\alpha};n_{\alpha},l_{\alpha};\alpha
\left\vert \widehat{H} \right\vert
\sigma_{\tilde{\alpha}},\lambda_{\tilde{\alpha}};n_{\tilde{\alpha}},l_{\tilde{\alpha}};
\tilde{\alpha}\right\rangle \right\Vert
\end{equation}
where
\begin{eqnarray}
& & \left\langle
\sigma_{\alpha},\lambda_{\alpha};n_{\alpha},l_{\alpha};\alpha
\left\vert
\widehat{H}\right\vert
\sigma_{\tilde{\alpha}},\lambda_{\tilde{\alpha}};n_{\tilde{\alpha}},l_{\tilde{\alpha}};
\tilde{\alpha}\right\rangle  \nonumber \\
&  = &\sum_{\nu_{\alpha}=1}^{N_{\alpha}} \sum_{\nu_{\tilde{\alpha}}=1}^{N_{\tilde{\alpha}}}
D_{\lambda_{\alpha},\nu_{\alpha}}^{\left(\sigma_{\alpha}, \alpha \right)}
\left\langle \nu_{\alpha},\lambda_{\alpha};
n_{\alpha},l_{\alpha}
\left\vert \widehat{H} \right\vert
\nu_{\tilde{\alpha}},\lambda_{\tilde{\alpha}};
n_{\tilde{\alpha}},l_{\tilde{\alpha}} \right\rangle
D_{\lambda_{\tilde{\alpha}},\nu_{\tilde{\alpha}}}^{\left(\sigma_{\tilde{\alpha}},
\tilde{\alpha} \right)}
\end{eqnarray}

This new representation exhibits the correct asymptotic behavior for large values of
$n_{\alpha}$ and $n_{\tilde{\alpha}}$, in the sense that off-diagonal matrix
elements coupling different channels, decrease to zero as
$n_{\alpha}$ and $n_{\tilde{\alpha}}$ tend to infinity.

Asymptotically the matrix has a tri-diagonal form from the kinetic energy of the
relative motion of the clusters. The diagonal matrix elements represent the interaction
within a given channel.

The third step in our approach consists of solving the set of equations
\begin{equation}
\sum_{c_{\tilde{\alpha}}}\sum_{n_{\tilde{\alpha}}=0}^{\infty}
\left\langle \sigma_{\alpha},\lambda_{\alpha};n_{\alpha},l_{\alpha};\alpha
\left\vert \widehat{H}-E \right\vert
\sigma_{\tilde{\alpha}},\lambda_{\tilde{\alpha}};n_{\tilde{\alpha}},l_{\tilde{\alpha}};
\tilde{\alpha}\right\rangle 
C_{n_{\tilde{\alpha}}l_{\tilde{\alpha}}}^{\left(c_{\tilde{\alpha}}\right)}=0
\label{eq:110}
\end{equation}
taking into account the appropriate boundary conditions to obtain either scattering or
bound state solutions.
This means that the solutions have to match the conditions (\ref{eq:104}) or
(\ref{eq:104a}) respectively, beyond some matching point $N_{i}$ that separates the
internal and asymptotic parts of the wave function. Thus e.g.\ for scattering we look
for solutions fo the form
\begin{eqnarray}
& & \left\{  C_{n_{\alpha}l_{\alpha}}^{\left(  c_{\alpha}\right)  }\right\}
=\left\{  C_{n_{\alpha}}^{\left(  c_{\alpha}\right)  }\right\}  =\left\{
C_{0}^{\left(  c_{\alpha}\right)  },C_{1}^{\left(  c_{\alpha}\right)  }%
,\ldots,C_{N_{i}}^{\left(  c_{\alpha}\right)  },\right. \nonumber\\
& & \left. \left\{ \sqrt{2r_{n_{\alpha}}}\left[  S_{c_{0},c_{\alpha}}%
\psi_{l_{\alpha}}^{\left(  -\right)  }\left(  p_{\alpha}r_{n_{\alpha}}\right)  -S_{c_{0},c_{\alpha}}\psi_{l_{\alpha}}^{\left(
+\right)  }\left(  p_{\alpha}r_{n_{\alpha}}\right)  \right]
;n_{\alpha}>N_{i}\right\} \right\} \label{eq:112}
\end{eqnarray}
where only the internal coefficients need to be determined. For simplicity we assume
that $N_{i}$ is identical for all channels. Inserting (\ref{eq:112}) into (\ref{eq:110})
then leads to
\begin{eqnarray}
& & \sum_{c_{\tilde{\alpha}}}
  \sum_{n_{\tilde{\alpha}}\leq N_{i}}
  \left\langle
  \sigma_{\alpha},\lambda_{\alpha};n_{\alpha},l_{\alpha}
  \left\vert \widehat{H}-E \right\vert
  \sigma_{\tilde{\alpha}},\lambda_{\tilde{\alpha}};n_{\tilde{\alpha}},l_{\tilde{\alpha}}
  \right\rangle C_{n_{\tilde{\alpha}}}^{\left(c_{\tilde{\alpha}}\right)} \nonumber \\
& & -\sum_{c_{\tilde{\alpha}}}S_{c_{0},c_{\tilde{\alpha}}}V_{c_{\alpha
,n_{\alpha};}c_{\tilde{\alpha}}}^{\left(  +\right)  }=-\sum_{c_{\tilde{\alpha
}}}\delta_{c_{0},c_{\tilde{\alpha}}}V_{c_{\alpha,n_{\alpha};}c_{\tilde{\alpha
}}}^{\left(  -\right)  }
\label{eq:113}
\end{eqnarray}
where e.g.
\begin{eqnarray}
&  V_{c_{\alpha,n_{\alpha};}c_{\tilde{\alpha}}}^{\left(  +\right)
}=V_{c_{\alpha,n_{\alpha};}c_{\tilde{\alpha}}}^{\left(  -\right)  \ast
}\nonumber\\
&  =\sum_{n_{\tilde{\alpha}}>N_{i}}\left\langle
\sigma_{\alpha},\lambda_{\alpha};n_{\alpha},l_{\alpha}\left\vert \widehat
{H}-E\right\vert \sigma_{\tilde{\alpha}},
\lambda_{\tilde{\alpha}};n_{\tilde{\alpha}},
l_{\tilde{\alpha}}\right\rangle \sqrt{2r_{n_{\tilde{\alpha}
}}}\psi_{l_{\tilde{\alpha}}}^{\left(  +\right)  }\left(  p_{\tilde{\alpha}%
}r_{n_{\tilde{\alpha}}}\right)  \label{eq:114}%
\end{eqnarray}
The solution of (\ref{eq:113}) then provides the explicit many-channel scattering wave function.
If the total number of binary channels is $N_{c}$, there are
$N_{c}\cdot N_{i}+N_{c}\cdot N_{c}$ equations for $N_{c}\cdot N_{i}$ expansion
coefficients of the internal part of the wave function, and $N_{c}\cdot N_{c}$
equations for the determination of the $S$-matrix.

\section{Results and discussion}

\subsection{Parameters of the calculation}

In the current calculations we consider a Minnesota nucleon-nucleon potential (MP)
for which we take the central part from \cite{kn:Minn_pot1}, and the spin-orbital part from
\cite{1970NuPhA.158..529R} (data set IV).
The exchange parameter $u$ is fixed at $u=0.956$ to reproduce the
relative positions of the $^{6}Li+p$ and $^{4}He+^{3}He$ thresholds.

To fix the oscillator bases we use a the same oscillator radius for both the $^{4}He$ and
deuteron clusters. We determine it by minimizing the energy of the three-cluster threshold
$^{4}He+d+p$, and obtain a value of $b=1.311$ fm. Table \ref{Tab:Thresholds}
shows a good agreement of the computed threshold energies of $^{6}Li+p$ and $^{4}He+d+p$
compared to experiment.
\begin{table}[htb!] \centering
\caption{$^{6}Li+p$ and $^{4}He+d+p$ threshold, w.r.t. the $^{4}He+^{3}He$
threshold (MeV).}
\label{Tab:Thresholds}\centering
\begin{tabular}
[c]{lrr}\hline
Threshold & MP & Experiment \cite{2002NuPhA.708....3T}\\\hline
$^{6}Li+p$ & $4.015$ & $4.020$\\
$^{4}He+d+p$ & $5.852$ & $5.493$\\\hline
\end{tabular}%
\end{table}%

To fix the set of Gaussian wave functions we follow the procedure of \cite{1994NuPhA.571..447V,1994PhRvC..50..189V}, and parametrize a set of widths $b_{\nu}$ with two variational
and parameters $a_{0}$ and $q$ as
\begin{equation}
b_{\nu}\ =a_{0}q^{\nu-1},\qquad\nu=1,2,\ldots
\end{equation}
This has been used in \cite{1994NuPhA.571..447V} and \cite{1994PhRvC..50..189V}
to obtain the ground state energy of $^{6}He$.

\subsection{Two-cluster subsystem properties}

\begin{figure}[htpb!]
\begin{center}
\includegraphics[width=0.75\textwidth]{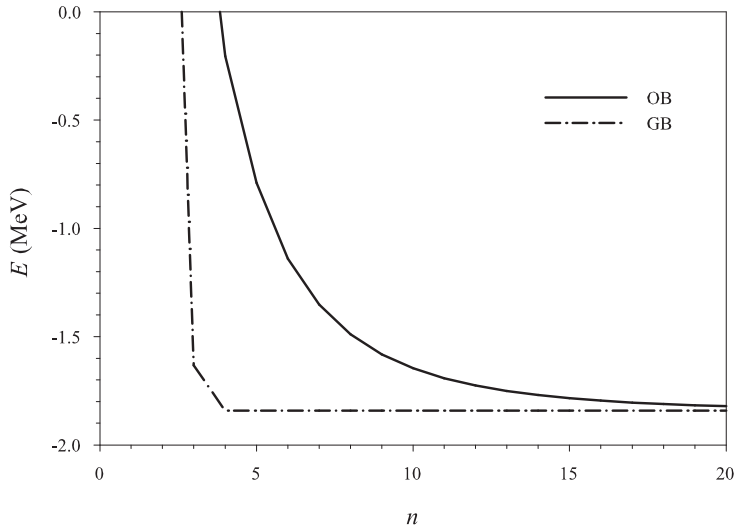}
\end{center}
\caption{$^{6}Li$ ground state energy with MP as a function of the
number of Gaussian (GB) and oscillator (OB) cluster states.}%
\label{fig:Li6_OB_GB}%
\end{figure}

We first elaborate on the merits of the Gaussian basis for the two-cluster subsystems.
To confirm its rapid convergence rate, we compare in Fig.\ \ref{fig:Li6_OB_GB}
the ground state energy of $^{6}Li$ for both the Gaussian basis ($a_{0}=1.0$ fm and $q=1.8$)
and the oscillator basis ($b=1.311$ fm). We have taken the latter value considered in this
paper, although it is not necessarily the optimal choice for this particular system.
One notices that convergence is reached with only 4 Gaussian functions, compared to more than
20 oscillator functions.

For the ground state of $^{3}He$ in the two-cluster $d+p$ model, a similar situation
occurs, and 4 Gaussian functions are sufficient with the $a_{0}=0.9$ fm, $q=1.8$ parametrization.

In Table \ref{Tab:PseudoBoundStates} we reproduce the energies of bound and
pseudo-bound states of $^{6}Li$ and $^{3}He$ obtained with four Gaussian functions.
\begin{table}[htbp] \centering
\caption{Spectrum of bound and pseudo-bound state of $^6Li$ and $^3He$  clusters.}
\begin{tabular}
[c]{rrrr}\hline
$\sigma$ & $^{3}He$ & $^{6}Li;L=0$ & $^{6}Li;L=2$\\\hline
$0$ & $-5.852$ & $-1.837$ & $2.463$\\
$1$ & $1.421$ & $3.308$ & $4.300$\\
$2$ & $8.765$ & $19.812$ & $15.307$\\
$3$ & $48.774$ & $77.898$ & $78.670$\\\hline
\end{tabular}
\label{Tab:PseudoBoundStates}
\end{table}

A standard Resonating Group Method (RGM) description takes one cluster function, i.e.\ a
single (oscillator) shell-model many-particle wave function, to describe the
internal structure of the interacting clusters. In this approximation the bound
state energy of $^{6}Li$ is 8.800 MeV,
and -3.018 MeV for $^{3}He$, which is way above the values obtained in Table
\ref{Tab:PseudoBoundStates}, as was to be expected from Fig.\ \ref{fig:Li6_OB_GB}.

To obtain stable results in the three-cluster model of $^{7}Be$ for its weakly
bound state, as well as for the elastic and inelastic scattering parameters, about
100 oscillator states must be considered in the calculation. This guarantees
the unitarity of the calculated $S$-matrix with high precision better than 0.1\%.

In a previous $^{4}He+^{3}He$ two-cluster model for $^{7}Be$ within a standard RGM approach
\cite{kn:cohstate2E}, \cite{2005PhRvC..71d4322S} stable bound and continuous
results were obtained with 30 to 50 oscillator states.

By taking into account cluster polarization, the GOB model allows for more
spatially dispersed clusters. This is confirmed by calculating the
root-mean-square-radius $R_{m}$ of the two interacting clusters. For $^{3}He$,
a one cluster function approach yields $R_{m}=1.311$ fm, while $R_{m}=1.696$ fm
with four Gaussian functions. For $^{6}Li$, these values are respectively
$R_{m}=1.650$ fm and $R_{m}=2.288$. It is therefore natural that more oscillator
states are necessary to properly reach the asymptotic region, because of the
relatively large distances between the clusters compared to their sizes.

\subsection{The Spectrum of $^{7}Be$}

\begin{figure}[hptb]
\begin{center}
\includegraphics[width=0.75\textwidth]{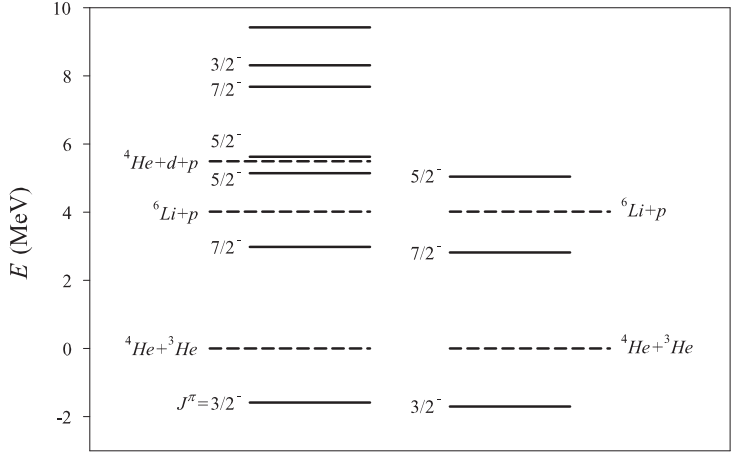}
\end{center}
\caption{Spectrum (relative to the $^{4}He+^{3}He$ threshold) of the ground and resonance
states of $^{7}Be$ in the GOB model with MP interaction. Theory (right) and experiment (left).}%
\label{Fig:Spectrum_MP_Exp}%
\end{figure}

In Table \ref{tab:7Be_spectrum_MP} we display the energy of the 3/2$^{-}$ (bound)
ground state, and the energies and widths of the resonance states of $^{7}Be$ obtained with
the MP interaction. All the energies are relative to the $^{4}He+^{3}He$ threshold.
\begin{table}[htbp!]
\centering
\caption{Ground state energy and resonance parameters $(E+i\Gamma)$ in
the GOB model of $^{7}Be$ with MP interaction (all in MeV and relative to the $^{4}He+^{3}He$
threshold).}
\label{tab:7Be_spectrum_MP}\centering
\begin{tabular}
[c]{lrr}\hline
State                    & Theory   & Experiment \cite{2002NuPhA.708....3T}\\\hline
$L=1$, $J^{\pi}=3/2^{-}$ & $-1.702$ & $-1.587$\\
$L=3$, $J^{\pi}=7/2^{-}$ & $2.820+i0.130$ & $\left( 2.983\pm0.05\right)
+i\left( 0.175\pm0.007\right)  $\\
$L=3$, $J^{\pi}=5/2^{-}$ & $5.040+i1.343$ & $\left(  5.143\pm0.10\right)
+i1.20$\\\hline
\end{tabular}%
\end{table}%
There is good agreement between theory and experiment for the
$3/2^{-}$ ground state. It is however slightly
overbound by 0.115 MeV, and the spin-orbital splitting energy is 0.16
MeV less than the experimental value. The energies and widths of the lowest
two resonances ($7/2^{-}$ and $5/2^{-}$) are very close to the experimental
value. The positions of the ground state and resonances of $^{7}Be$ are displayed
in Fig.\ \ref{Fig:Spectrum_MP_Exp}.

We now turn to the effect of the polarization of the two-cluster subsystems
$^{6}Li$ and $^{3}He$ on the ground state energy of $^{7}Be$. We do so comparing results
with ( marked ''Y'') and without (marked ''N'') polarization of the subsystem in Table
\ref{Tab:Polariz_GrounSt}. We suppress the polarization by using only a single function
instead of all eigenfunctions of the corresponding two-cluster hamiltonian in the calculations.
This corresponds to a rigid cluster throughout the calculation, whereas using the full
set of eigenfunctions allows for adapting the size and shape of the subsystem to the
presence of the third cluster.
One notices that the polarization of $^{6}Li$ has a stronger impact than that
of $^{3}He$.
\begin{table}[htbp]
\centering
\caption{Polarization (Y: included, N: suppressed) effect on the $^7Be$ ground state energy.} \label{Tab:Polariz_GrounSt}\centering
\begin{tabular}
[c]{ccr}\hline
$^{3}He$ & $^{6}Li$ & $E$(MeV)\\\hline
N & N & -0.971\\
Y & N & -1.413\\
N & Y & -1.666\\
Y & Y & -1.702\\\hline
\end{tabular}
\end{table}

\begin{figure}[hptb]
\begin{center}
\includegraphics[width=0.75\textwidth]{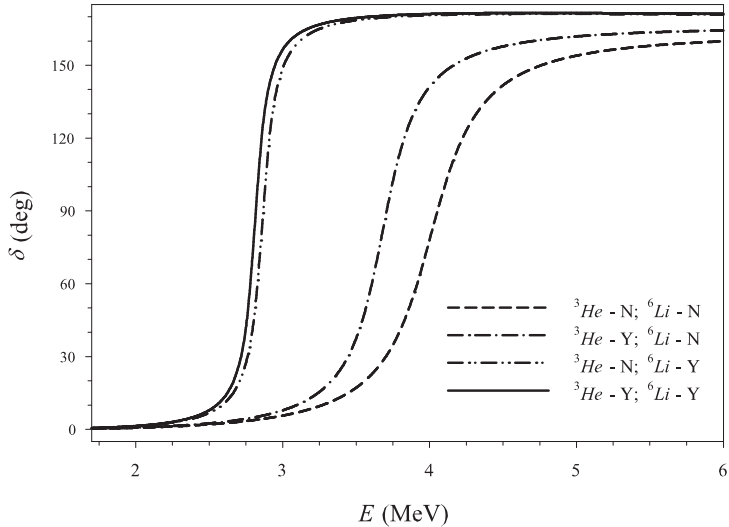}
\end{center}
\caption{Polarization effects on the phase shift of $^{4}He+^{3}He$ elastic
scattering.}
\label{Fig:Phases_4He_3He_72_Pol}
\end{figure}

Using the same (Y,N) approach we look at the effect of cluster polarization on the phase
shift of $^{4}He+^{3}He$ elastic scattering with total momentum $J^{\pi}=7/2^{-}$ ($L=3$)
in Fig.\ \ref{Fig:Phases_4He_3He_72_Pol}.
Similarly, in table \ref{Tab:Polariz_Resonances} we look at the effect on the parameters
for the 7/2$^{-}$ and 5/2$^{-}$ resonance states.

The cluster polarization substantially decreases the resonance energy and width, which
points to an increase in the effective interaction between the clusters. The same
observation has been made in \cite{kn:VVS+1984MonResE,Deumens84,kn:VVS+1984Reson8E,kn:cohstate2E, kn:VVS+1990RGM+Sp2RE,kn:VV86collresE,2005PhRvC..71d4322S,kn:cohstate1E} for
collective monopole and quadrupole polarizations. One again notices that the
polarization of the $^{6}Li$ cluster is more important than that of the of
$^{3}He$ cluster. The resonance properties, obtained with $^{6}Li$
polarization are only marginally different from those calculated with both $^{6}Li$
and $^{3}He$ polarizations.
\begin{table}[htbp]
\centering
\caption{Polarization effects on the resonance properties of $^{7}Be$.}
\label{Tab:Polariz_Resonances}\centering
\begin{tabular}
[c]{ccll}\hline
&  & $J^{\pi}=7/2^{-}$ & $J^{\pi}=5/2^{-}$\\\hline
$^{3}He$ & $^{6}Li$ & $E+i\Gamma$(MeV) & $E+i\Gamma$(MeV)\\\hline
N & N & $4.01+i0.51$ & $5.86+i2.42$\\
Y & N & $3.69+i0.37$ & $5.74+i2.24$\\
N & Y & $2.87+i0.14$ & $5.08+i1.38$\\
Y & Y & $2.82+i0.13$ & $5.04+i1.34$\\\hline
\end{tabular}
\end{table}

\subsection{$^{7}Be$ ground state properties}

The electromagnetic observables can calculated using the explicit ground state wave function.
In Table \ref{Tab:Radii&Q} we list root-mean-square radii (proton ($R_{p}$), neutron ($R_{n}$)
and mass ($R_{m}$)) and the quadrupole moments (proton ($Q_{p}$), neutron ($Q_{n}$) and mass
($Q_{m}$)) for the 3/2$^{-}$ ground state. As one expects, the proton radius is larger than
the neutron one. The quadrupole moment is an indicator of the deformation of the nucleus, and
a negative value corresponds to a prolate deformation. The differences in these moments reflect
the pronounced cluster structure of the $^{7}Be$ ground state.
\begin{table}[htbp] \centering
\caption{Radii and quadrupole moments of the $^{7}Be$ 3/2$^{-}$ ground state.}
\begin{tabular}
[c]{lrrrrrr}\hline
$J^{\pi}$ & GOB & MCRGM\cite{2002NuPhA.699..963A} & SCRGM\cite{1984NuPhA.413..323K}
 & SVM\cite{1997NuPhA.616..383V} & SM\cite{2006PhLB..634..191N} & Experiment\\\hline
$R_{p}$ (fm) & 2.457 &  & 2.74 & 2.41 & 2.342 & 2.53$\pm0.03~$%
\cite{1985PhRvL..55.2676T}\\
$R_{n}$ (fm) & 2.263 &  & 2.50 & 2.31 &  & \\
$R_{m}$ (fm) & 2.375 &  &  & 2.36 &  & \\
$Q_{p}$ ($e\cdot$fm$^{2}$) & -6.245 & -6.4 & -6.125 & -6.11 & -5.153 & \\
$Q_{n}$ ($e\cdot$fm$^{2}$) & -3.739 &  &  &  &  & \\
$Q_{m}$ ($e\cdot$fm$^{2}$) & -9.984 &  &  &  &  & \\\hline
\end{tabular}
\label{Tab:Radii&Q}
\end{table}

In Table \ref{Tab:Radii&Q} we compare our GOB results with those of the multi-channel RGM of
Arai et al (MCRGM) \cite{2002NuPhA.699..963A}), of the single-channel RGM of Kajino et al.\
(SCRGM) \cite{1984NuPhA.413..323K}), of the Stochastic Variational Method (SVM)
\cite{1997NuPhA.616..383V} and of the Shell Model (SM) calculations \cite{2006PhLB..634..191N}.

The GOB results are close to those of SVM
\cite{1997NuPhA.616..383V}, and of the multi-channel RGM \cite{2002NuPhA.699..963A}.
These results have also obtained with the MP interaction.

The single-channel SCRGM results of
\cite{1984NuPhA.413..323K} have been obtained with the modified
Hasegawa-Nagata potential. The ground state quadrupole moment is comparable,
whereas the charge radius is much larger in the SCRGM.

The Shell Model calculation of
\cite{2006PhLB..634..191N} has been performed with the Bonn nucleon-nucleon potential. Its
results for the charge radius and quadrupole moments are smaller than in the cluster models
(MCRGM, SCRGM, SVM,GOB). This can be attributed to the fact that the
Shell Model, involving a $10\hbar\Omega$ or $12\hbar\Omega$ state space, confines the
inter-cluster distances in the $^{7}Be$ nucleus. In comparison,
the GOB model uses a model space of $200\hbar\Omega$ for
the inter-cluster behavior of the dominant $^{4}He+^{3}He$ cluster configuration.

\subsection{Comparing polarization methods}
As discussed in the introduction, one can distinguish collective polarization
(quadrupole and monopole) for the compound nucleus as a whole (QM), RGM models with
monopole excitations of the individual clusters (MRGM), and cluster polarization as described
in the GOB model. It is interesting to compare these different polarization methods
to gauge their impact on the ground state energy.

A detailed study in MRGM is only available for $^7Li$ \cite{1986PhRvC..34..771K}, so we
investigate the three methods applied to this nucleus, as it should have deformation
properties comparable to $^7Be$.

To make the comparison consistent, we use the MP interaction parameters of
\cite{1986PhRvC..34..771K}, i.e.\ $u=1.0174$ and a spin-orbital strength of 0.821 (it is 1.0
for the other calculations in this paper, as suggested in
\cite{1970NuPhA.158..529R}). With this interaction, we have performed
the QM calculations along the lines reported in \cite{2006JPhG...32.2137S} and
\cite{2005PhRvC..71d4322S}, a GOB calculations for a single $\alpha+t$ channel (SGOB)
and a full GOB calculation as outlined in this paper. A standard RGM calculation without
cluster polarization (SRGM) has been included as a benchmark.

\begin{table}[htbp] \centering%
\caption{Ground state energy of $^7Li$ without polarization
(Standard RGM) and with different types of polarization (see text)}
\begin{tabular}{rrrrr}
\hline
SRGM    & MRGM    & SGOB    & QM      & GOB \\ \hline
-31.465 & -32.027 & -33.227 & -33.777 &  -34.150 \\ \hline
\end{tabular}
\label%
{Tab:Spectr7Li3Pol}%
\end{table}

In Table \ref{Tab:Spectr7Li3Pol} the results for the different polarization
models are displayed. One notices that the collective QM and SGOB models, which
are comparable in terms of model space, improve with respect to the MRGM that
includes only monopole cluster polarization. Clearly the full GOB calculation (including
cluster polarization of $^6Li$ in the $^6Li+n$ channel) provides
the best results. This was to be expected from the above $^7Be$ results.

We notice that in the MRGM calculation, the monopole polarization
of $^3He$ is obtained by using four Gaussian functions resulting in -5.906 MeV
of binding energy. In the SGOB model four Gaussian cluster functions are used
to describe cluster polarization of $^3He$, and yields a binding energy of -5.953 MeV.
This confirms the above results.

We can thus state that the cluster polarization of the channel subsystems described in the
GOB model indeed plays an important role in seven-nucleon systems, and currently represents
most prominent polarization type.

\subsection{Three-cluster geometry}

The expansion coefficients  $\left\{C_{\nu_{\alpha}\lambda_{\alpha};n_{\alpha}l_{\alpha}}^{\left(\alpha\right)}\right\}$
determine the three-cluster wave function $\Psi^{J}$ (\ref{eq:003b}) and also the Faddeev
components $f_{\alpha}\left(\mathbf{x}_{\alpha},\mathbf{y}_{\alpha}\right)$
($\alpha=1,2,3$) (\ref{eq:003}). We use the latter representation to analyze the wave function.

The correlation function for the ground state is
\begin{equation}
P_{\alpha}\left(  x_{\alpha},y_{\alpha}\right)  =x_{\alpha}^{2}y_{\alpha}%
^{2}\int\left\vert f_{\alpha}\left(  \mathbf{x}_{\alpha},\mathbf{y}_{\alpha
}\right)  \right\vert ^{2}d\widehat{\mathbf{x}}_{\alpha}d\widehat{\mathbf{y}%
}_{\alpha} \label{eq:201}%
\end{equation}
where the integration runs over the unit vectors
$\widehat{\mathbf{x}}_{\alpha}$ and $\widehat{\mathbf{y}}_{\alpha}$.
\begin{figure}[htb!]
\begin{center}
\includegraphics[width=0.6\textwidth]%
{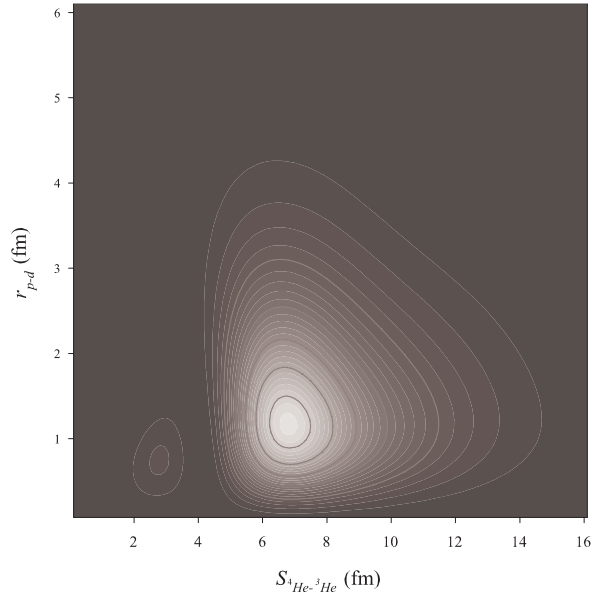}%
\caption{Correlation function for the $^{4}He+^{3}He$ binary channel.}%
\label{Fig:Corr_fun_1}%
\end{center}
\end{figure}

To interpret the the polarizability in terms of the three-cluster structure, we introduce
the root-mean-square radii $R_{\alpha}$:
\begin{equation}
R_{\alpha}\left(  y_{\alpha}\right)  =\sqrt{\int d\widehat{\mathbf{y}}%
_{\alpha}\int x_{\alpha}^{2}\left\vert f_{\alpha}\left(  \mathbf{x}_{\alpha
},\mathbf{y}_{\alpha}\right)  \right\vert ^{2}d\mathbf{x}_{\alpha}%
/\mathcal{N}_{\alpha}\left(  y_{\alpha}\right)  } \label{eq:202}%
\end{equation}
and
\begin{equation}
\mathcal{N}_{\alpha}\left(  y_{\alpha}\right)  =\int d\widehat{\mathbf{y}%
}_{\alpha}\int\left\vert f_{\alpha}\left(  \mathbf{x}_{\alpha},\mathbf{y}%
_{\alpha}\right)  \right\vert ^{2}d\mathbf{x}_{\alpha}%
\end{equation}
They represent the root-mean-square radius of a two-cluster subsystem as a function of its
distance to the third cluster (i.e.\ the distance of the centre-of-mass of the two-cluster
system to the centre-of-mass of the third cluster).
\begin{figure}[htb!]
\begin{center}
\includegraphics[width=0.6\textwidth]%
{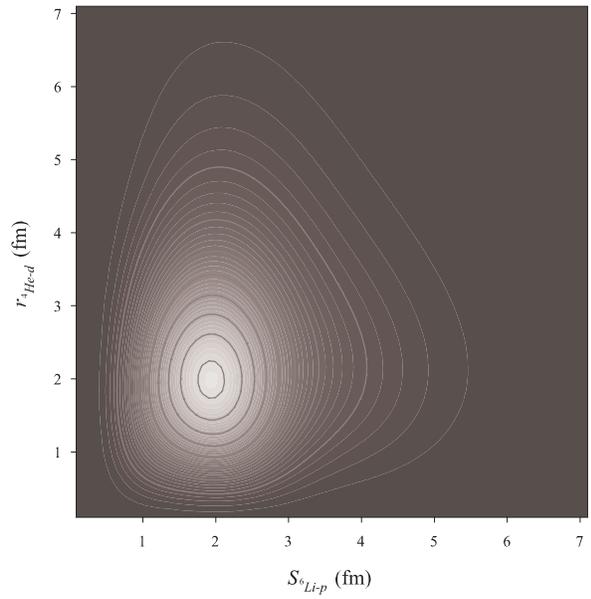}%
\caption{Correlation function for the $^{6}Li+p$ binary channel}%
\label{Fig:Corr_fun_2}%
\end{center}
\end{figure}

We introduce new coordinates $\mathbf{r}_{\alpha}$ and $\mathbf{S}_{\alpha}$
whose norms ($r_{\alpha}$ and $S_{\alpha}$ correspond to the distances between the centers
of mass of the clusters, that allow for a proper interpretation of the quantities
(\ref{eq:201}) and (\ref{eq:202}).
They relate to the original Jacobi coordinates as:
\begin{equation}
\mathbf{x}_{\alpha}=\sqrt{\frac{A_{\beta}A_{\gamma}}{A_{\beta}+A_{\gamma}}
}\mathbf{r}_{\alpha},\quad\mathbf{y}_{\alpha}=\sqrt{\frac{A_{\alpha}\left(
A_{\beta}+A_{\gamma}\right)  }{A_{\alpha}+A_{\beta}+A_{\gamma}}}
\mathbf{S}_{\alpha}
\end{equation}

Fig.\ \ref{Fig:Corr_fun_1} displays the correlation function for the
$^{4}He+^{3}He$ channel and Fig.\ \ref{Fig:Corr_fun_2} for the $^{6}Li+p$ channel.
One notices from Fig.\ \ref{Fig:Corr_fun_1} that the distance between $^{4}He$ and
$^{3}He$ is approximately 8 fm, and much larger than the separation of $d$ and $p$.
From Fig.\ \ref{Fig:Corr_fun_2} one observes that the binary cluster
configuration $^{6}Li+p$ is surprisingly compact. Both the distance between $^{4}He$
and $d$ and between $^{6}Li$ and $p$ are around 1 fm.
These different geometric configurations can be related to the
ground state energy relative to the thresholds of the corresponding binary channel.
Indeed, the ground state is positioned at -$1.702$ MeV
from the $^{4}He+^{3}He$ threshold and at -$5.722$ MeV with from the $^{6}Li+p$
threshold. This agrees with a very dispersed $^{4}He+^{3}He$
and a compact $^{6}Li+p$ configuration.
\begin{figure}[htb!]
\begin{center}
\includegraphics[width=0.75\textwidth]
{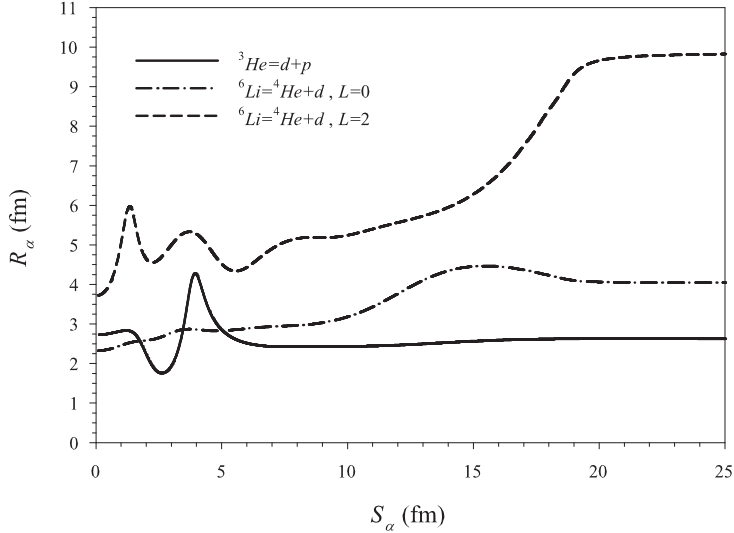}
\caption{Dependence of the root-mean square radius $R_{\alpha}$ of the two-cluster
subsystems on the distance $S_{\alpha}$ from the third cluster.}
\label{Fig:RMSR_pol}
\end{center}
\end{figure}

In Fig.\ \ref{Fig:RMSR_pol} we show the dependence of the root-mean-square radius
of the two-cluster subsystems on the distance from the third cluster. One notices
how $^{6}Li$ strongly adapts its size when the proton is at large distances (more
than 15 fm) from $^{6}Li$.
When the $^{6}Li$ and the proton are near,the $^{6}Li$ $0^{+}$ ground state gets
compressed approximately 1.5 times, and the $2^{+}$ excited state approximately two times.
The $^{3}He$ nucleus, in a two-cluster configuration $d+p$, is strongly affected when
the $\alpha$-particle is closer than 6 to 7 fm. Note that without polarization all
three curves in Fig.\ \ref{Fig:RMSR_pol} would be horizontal lines, as
the two-cluster subsystems then have a constant size. The figure illustrates the impact
of cluster polarization of $^{6}Li$ and $^{3}He$ in the description of $^{7}Be$.
\begin{figure}[htb!]
\begin{center}
\includegraphics[width=0.75\textwidth]{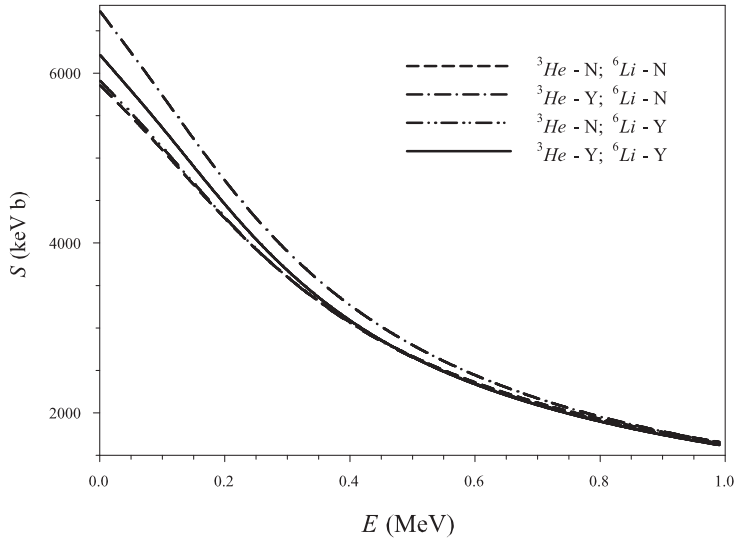}
\end{center}
\caption{Astrophysical $S$-factor of the reaction $^{6}Li(p,^{3}He)^{4}He$ for the dominant
$L=0$ component with and without $^{3}He$ and $^{6}Li$ polarization.}%
\label{Fig:S_factor_pol}%
\end{figure}

\subsection{The Reaction $^{6}Li\left(  p,^{3}He\right)  ^{4}He$}

The astrophysical $S$-factor of the reaction $^{6}Li\left(
p,^{3}He\right)^{4}He$ is obtained from the total cross section, which has been computed
with four total angular momentum $L=0$, $1$, $2$ and $3$. We have observed that zero angular
momentum contribution dominates the low energy range $0\leq E\leq1$ MeV of the cross section.

Fig.\ \ref{Fig:S_factor_pol} displays the $S$-factor of the 
$^{6}Li\left(  p,^{3}He\right)  ^{4}He$ reaction calculated for $L=0$.
In contrast to the analysis of the ground state and resonance energies,
the effects of cluster polarization are less evident here. In Fig.\ \ref{Fig:S_factor_pol} it
is seen that $^{3}He$ polarization increases the $S$-factor, thus increasing the coupling
between the channels. The $^{6}Li$ polarization only has a small effect on the $S$-factor.

\begin{figure}
[hptb]
\begin{center}
\includegraphics[width=0.75\textwidth]%
{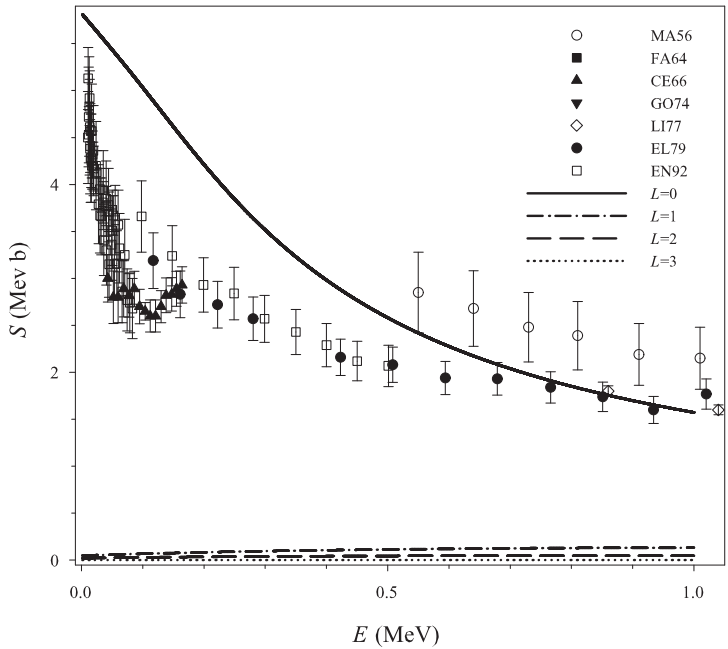}%
\caption{$S$-factor of the reaction $^{6}Li\left(  p,^{3}He\right)  ^{4}He$.
Experimental data are taken from \cite{MA56}- MA56, \cite{FA64}-FA64,
\cite{GE66}-GE66, \cite{GO74}-GO74, \cite{LI77}-LI77, \cite{EL79}-EL79,
\cite{EN92}-EN92.}%
\label{Fig:SFactor_T&S}%
\end{center}
\end{figure}

In Fig.\ \ref{Fig:SFactor_T&S} we compare the results for the $S$-factor with
both $^{6}Li$ and $^{3}He$ cluster polarization with the available
experimental data \cite{MA56,FA64,GE66,FI67,SP71,GO74,LI77,EL79,EN92}. Notations and
data are taken from \cite{1999NuPhA.656....3A} and the web site
http://pntpm.ulb.ac.be.
One observes that our model somewhat overestimates the $S$-factor in the
low energy range \cite{LI77}. A similar result was obtained in
\cite{2002NuPhA.699..963A}. A possible reason is the lack
of tensor components in the MP interaction. These would couple
channels with different total spin and total angular momentum,
and thus reduce the coupling between entrance and exit channels.

\section{Conclusions}

We have introduced a fully microscopic three-cluster model in which an expansion
in terms of Faddeev components is used. The set of equations for the
Faddeev amplitudes is derived and solved within the Coupled Reaction Channel
Formalism.
The dynamics of the three-cluster system is described by an effective
two-body nucleon-nucleon potential, and takes into account the Pauli exchange
principle correctly.

We have used two different expansion bases: (1) a Gaussian basis suitable for
the description for the different two-cluster subsystems, and (2) an oscillator
basis to incorporate the appropriate boundary conditions for bound and
continuous states in a many-channel compound system. 
Only few Gaussian states are required to describe the ground state of the
(bound) two-cluster subsystems, such as $^{6}Li$ which has a high degree
of $\alpha + d$ clusterization. This limits the computational effort involved
in the calculations.
As for the three-cluster results, many more oscillator states (typically 70 to 100)
are needed to guarantee both the convergence of the ground state, and the
unitarity of the many channel $S$-matrix to sufficient precision.

The method has been applied to investigate cluster polarizability in the
ground and resonance states of $^{7}Be$ for which the three-cluster configuration
$^4He+d+p$ was used. This provides for (i) the two binary channels $^3He+^4He$
and $^6Li+p$, which are prominent in the low energy region of $^7Be$, and (ii) two
bound two-cluster subsystems ($^6Li$ as $^4He+d$, $^3He$ as $d+p$). The latter are
modeled with a Gaussian basis expansion, and allow to
study the relative behavior of the containing clusters when $^6Li$ (respectively $^3He$)
collides with a proton (respectively an $\alpha$-particle). We refer to this behavior
as ``cluster polarization''. The inclusion of cluster polarization in $^7Be$ leads
to a strong decrease of the energy of the ground and resonance states, and
reduces the resonance widths two to four times. The $^6Li$ polarization is
more important in this respect than the $^3He$ one.

The effect of cluster polarizability was also studied in the reaction 
$^{6}Li\left(p,^{3}He\right) ^{4}He$ by considering the $S$-factor, for which
the $^3He$ polarization was seen to be more important.

\section{Acknowledgments}

Support from the Fonds voor Wetenschappelijk Onderzoek Vlaanderen (FWO),
G0120-08N is gratefully acknowledged. V.~S.~Vasilevsky is grateful to the
Department of Mathematics and Computer Science of the University of Antwerp
(UA) for hospitality.
This work was supported in part by the Program of Fundamental Research of the
Physics and Astronomy Department of the National Academy of Sciences of Ukraine.
\bibliographystyle{00}

%
%
%
%
%
%
%

\end{document}